\documentclass[12pt,preprint]{aastex}
\usepackage{graphics}

\def\be{\begin{equation}}
\def\ee{\end{equation}}
\def\bq{\begin{eqnarray}}
\def\eq{\end{eqnarray}}

\begin{document}

\title{Gamma Ray Bursts from delayed collapse \\of neutron stars 
       to quark matter  stars}

\author{Z.~Berezhiani\altaffilmark{1}, 
I.~Bombaci\altaffilmark{2}, 
A.~Drago\altaffilmark{3}, 
F.~Frontera\altaffilmark{3,4} and
A.~Lavagno\altaffilmark{5}}

\altaffiltext{1}{Dip. di Fisica, Univ. di L'Aquila and INFN
Sez. del Gran Sasso, I-67010 Coppito, Italy}
\altaffiltext{2}{Dip. di Fisica, Univ. di Pisa 
and INFN Sez. di Pisa, I-56127 Pisa, Italy}
\altaffiltext{3}{Dip. di Fisica, Univ. di Ferrara and INFN
Sez. di Ferrara, I-44100 Ferrara, Italy}
\altaffiltext{4}{IASF, CNR, I-40129 Bologna, Italy}
\altaffiltext{5}{Dip. di Fisica, Politecnico di Torino and INFN
Sez. di Torino, I-10129 Torino, Italy}

\begin{abstract}
We propose a model to explain how a Gamma Rays Burst can take place days or years after a 
supernova explosion.
Our model is based on the conversion of a pure hadronic star (neutron star) into a star
made at least in part of deconfined quark matter. The conversion process can be delayed if
the surface tension at the interface between hadronic and deconfined-quark-matter
phases is taken into account.
The nucleation time ({\it i.e.} the time to form a critical-size drop of 
quark matter) can be extremely long if the mass of the star is small.  
Via mass accretion the nucleation time can be dramaticaly reduced 
and the star is finally converted into the stable configuration.   
A huge amount of energy, of the order of 10$^{52}$--10$^{53}$ erg, 
is released during the conversion process and can produce a powerful Gamma Ray Burst. 
The delay between the supernova explosion generating the metastable  
neutron star and the new collapse can explain the delay proposed in
GRB990705 \citep{Amati00} 
and in GRB011211 \citep{Reeves02}.
\end{abstract}


\vspace{1.5cm}
\noindent

\section{Introduction}
The discovery of a transient (13~s) absorption feature in the prompt 
emission of the $\sim 40$~s Gamma Ray
Burst (GRB) of July 5, 1999 (GRB990705) \citep{Amati00} and the evidence 
of emission features in the afterglow of several GRBs
\citep{Piro99,Yoshida99,Piro00,Antonelli00,Reeves02} have stimulated the interpretation
of these characteristics in the context of the fireball model of GRBs. 
\citet{Amati00} attribute the transient absorption feature
of GRB990705 (energy released $\sim 10^{53}$ erg assuming isotropy) to a redshifted 
K edge of Iron contained in an environment not far 
from the GRB site ($\sim 0.1$~pc) and crossed by the GRB emission. 
They estimate an Iron abundance
typical of a supernova (SN) environment ($A_\mathrm{Fe} \sim 75$) and a time delay of 
about 10 years between the SN explosion and the GRB event. \citet{Lazzati01}
give a different interpretation
of the absorption feature, in terms of a redshifted resonance scattering feature 
of H--like Iron (transition 1s--2p, $E_\mathrm{rest} =6.927$~keV) in an inhomogeneous 
high--velocity outflow, but invoke a Iron rich environment as well, due to a preceding SN 
explosion, even if a shorter time delay ($\sim 1$~yr) between SN and GRB is inferred. 
A SN explosion preceding  the GRB event is also inferred for explaining the properties 
of the emission features in the X--ray afterglow spectrum of GRB000214 \citep{Antonelli00} 
and GRB991216 \citep{Piro00}. 
In the latter case it cannot be excluded that the SN explosion occured 
days or weeks before the GRB \citep{Rees00}.
\citet{Reeves02}, to explain the multiple
emission features observed in the afterglow spectrum of GRB011211
(time duration of $\sim 270$~s, isotropic gamma--ray energy of $5 \times 10^{52}$~erg), 
invoke a SN explosion preceding the GRB event by $\sim 4$~days (in the isotropic limit,
a minimum of 10 hrs). Even if other interpretations for the afterglow 
emission lines are possible which do not involve a previous SN explosion 
(e.g., \citep{Rees00,Meszaros01}), 
this explosion seems to be the most likely way to explain the transient 
absorption line observed from GRB990705 \citep{Bottcher02}. 
In conclusion, the previous observations suggest that, at least for a certain number of GRBs,
a SN explosion happened before the GRB, with a time interval between the two events ranging
from a few hours to a few years.
In this context, 
an attractive scenario is that described by the {\it supranova} model \citep{Vietri98} 
for GRBs. In this model, the GRB is the result of the collapse to a black hole (BH) of a
supramassive fast rotating neutron star (NS), as it loses angular momentum.   
According to this model the NS is produced in the SN explosion preceding the GRB event. 
The initial barionic mass $M_B$ of the NS is assumed to be above the maximum baryonic mass 
for non-rotating configurations. However, as also noticed by \citet{Bottcher02}, 
on the basis of realistic calculations of collapsing NS \citep{fryer1998}, 
in these collapses too much baryonic material is ejected and thus the energy output 
is expected to be too small to produce GRBs. Even if the introduction of magnetic fields 
or beaming could overcome this limitation, in any case, the GRB duration from a NS 
collapse should be very short ($\ll1$~s), much  shorter than that observed from 
GRB990705. 

In this Letter, we propose an alternative model to explain the existence
of GRBs associated with previous SN explosions.  
In this model, unlike the supranova model, the NS collapse to BH
is replaced by the conversion from a metastable, purely hadronic star (neutron star) 
into a more compact star in which deconfined quark matter (QM) is present. 
This possibility has already been discussed in the literature \citep{cd96,bd00,wang,ouyed}.
The new and crucial idea we introduce here, is the metastability  
of the purely hadronic star due to the existence of a non-vanishing 
surface tension at the interface separating hadronic matter
from quark matter. The mean-life time of the metastable NS can then
be connected to the delay between the supernova explosion and the GRB. 
As we shall see, in our model we can easily obtain a burst lasting tens 
seconds, in agreement with the observations.
The order of magnitude of the energy released is also the appropriate one.

\section{Quark Matter nucleation in compact stars}
Recently various possibilities have been discussed in the literature to get   
compact stars in which matter is, partially or totally, in a state of deconfined 
quarks (see {\it e.g.} \citet{glenbook,martino}; \citet{andrea}).  
Concerning the stellar quark content, it is possible to have three different classes 
of compact stars: a) purely hadronic stars (HS), in which no fraction of QM is 
present; b) hybrid stars (HyS), in which only at the center of the star QM is 
present either as a mixed phase of deconfined quarks and hadrons or as a pure phase; 
c)quark stars (QS), in which the surface of the star is made of matter 
having a large density, of the order of nuclear matter saturation density or larger, 
and the bulk of the star is made of deconfined QM.
The sizeable amount of observational data collected by the new generations 
of X-ray satellites, has provided a growing body of evidence for the 
existence of very compact stars, which could be
HyS or QS \citep{bomb97,che98,li99a,li99b,xu02,dra02}.

In our scenario, we consider a purely HS whose central density 
(pressure) is increasing due to spin-down or due to mass accretion 
(e.g., from fallback of ejected material in the SN explosion).  
As the central density approaches the deconfinement critical density, 
a virtual drop of quark matter can be formed in the central region of the star.
The fluctuations of a spherical droplet of quark matter having a radius $R$ 
are regulated by a potential energy of the form \citep{lif}
\be
U(R)={4 \over 3} \pi R^3 n_q (\mu_q-\mu_h)+4 \pi\sigma R^2
+ 8 \pi\gamma R
\ee
where $n_q$ is the quark baryon density, $\mu_h$ and $\mu_q$ are 
the hadronic and quark chemical potentials at a fixed pressure $P$,
and $\sigma$ is the surface tension for the surface separating
quarks from hadrons. Finally, the term containing $\gamma$ is the 
so called curvature energy. 
The value of the surface tension $\sigma$ is poorly known, 
and typical values used in the literature range from
10 to 50 MeV/fm$^2$ \citep{hei,iida}.  
Following the work of \citet{iida}, we have 
assumed that the term with $\sigma$ takes into account 
in an effective way also the curvature energy. The term with $\gamma$ is discussed {\it e.g.} 
by \citet{madsen}, while other more complicated terms, connected with the Coulomb energy, are
discussed in the literature \citep{hei,iida}. We have neglected 
them in our analysis since they do not dramatically modify
both the nucleation time and the energy associated with the 
transition into the stable quark matter configuration.  

If the temperature is low enough, the process of formation of a bubble having a 
critical radius proceeds through quantum tunnelling and it 
can be computed using a semiclassical approximation. The procedure
is rather straightforward. First one computes, 
using the semiclassical (WKB) approximation, the ground state energy 
$E_0$ and the oscillation frequency $\nu_0$ of the virtual QM drop 
in the potential well $U(R)$. Then it is possible to calculate in a relativistic frame
the probability of tunneling as \citep{iida}:
\be
p_0=\exp [-{A(E_0)\over \hbar}]
\ee
where
\be
A(E)={2}\int_{R_-}^{R_+} dR \sqrt{[2 M(R)+E-U(R)][U(R)-E]}\, .
\ee
Here $R_\pm$ are the classical turning points and 
\be
M(R)=4\pi\rho_h(1-{n_q\over n_h})^2 R^3 \,\,\,\,\, ,
\ee
$\rho_h$ being the hadronic energy density (here and in the following we adopt
the so-called ``natural units'', in which $\hbar=c=1$).
$n_h$ and $n_q$ are the baryonic densities at a same and given
pressure in the hadronic and quark phase, respectively. The nucleation time is then equal to 
\be
\tau = (\nu_0 p_0 N_c)^{-1}\, ,
\ee
where $N_c$ is the number of centers of droplet formation in the star, and it is 
of the order of $10^{48}$ \citep{iida}.

\section{Results}
The typical mass-radius relations for the three types
of stars we are discussing can be found e.g. in Fig. 3 of \citet{andrea},
where a relativistic non-linear Walecka-type model \citep{gm} has been used
to describe the hadronic phase. As it appears,
stars containing QM (either HyS or QS) are more compact than
purely HS. In particular QS can have much smaller
radii then HS when they have a small mass. In our scenario a metastable
HS having a mass of, e.g., 1.3 $ M_\odot$ and a radius of 
$\sim 13$ km can collapse into an HyS having a radius of $\sim 10.5$ km
or into a QS with radius $\sim 9$ km 
(with respect to the results of Fig. 3 of \citet{andrea}, here we use an equation of 
state (EOS) which includes hyperonic degrees of freedom).  
The nature of the stable configuration reached after the stellar conversion 
({\it i.e.} an HyS or a QS) will depend on the parameters of the quark phase EOS.    

The time needed to form a critical droplet of deconfined quark matter 
can be calculated for different values of the stellar central pressure $P_c$ 
(which enters in the expression of the energy barrier in eq. (1)) 
and it can be plotted as a function of the gravitational mass $M_\mathrm{HS}$  
of the HS corresponding to that given value of central pressure. 
The results of our calculations for a specific EOS of hadronic matter 
(the GM3 model with hyperons of \citet{gm}) are reported in Fig. 1, 
where each curve refers to a different value of the bag constant $B$. 
If we assume, for example, $B^{1/4}=170$ MeV (which corresponds to
$B=109$ MeV/fm$^3$) and  the initial mass 
of the HS to be $M_{HS} = 1.32~ M_\odot$, we find that the ``life time'' 
for this star is about $10^{12}$ years.  
As the star accretes a small amount of matter, the consequential increase 
of the central pressure lead to a huge reduction of the nucleation time, 
and, as a result, to a dramatic reduction of the HS life time. 
For our HS with initial mass of $1.32~M_\odot$ the accretion of about 
$0.01~ M_\odot$ reduces the star life time to a few years.  
We would like to stress that in our model the delay between the SN explosion
and the GRB is regulated by the mass accretion rate, rather then by
the mass and the spinning of the metastable star itself. Since the mass
accretion rate is generally larger during the first days after the SN explosion,
a delay of a few days will be rather typical in our scenario. However, longer
delays are also possible if the material ejected during the SN explosion has a small
fallback.

In the model we are presenting, the GRB is due to the cooling of the
justly formed HyS or QS via neutrino-antineutrino emission (and maybe
also via emission of axion-like particles, see below). The subsequent 
neutrino-antineutrino annihilation generates the GRB.
In our scenario the duration of the prompt emission of the GRB is therefore
regulated by two mechanisms: 1) the time needed for the conversion
of the HS into a HyS or QS, once a critical-size droplet is formed and 
2) the cooling time of the justly formed HyS or QS.
Concerning the time needed for the conversion into QM of at
 least a fraction of the star, the seminal work by
\citet{Oli87} has been reconsidered by \citet{HB88}. The conclusion
of this latter work is that the stellar conversion is a very fast process,
having a duration much shorter than 1s. On the other hand, 
the neutrino trapping time, which provides the cooling time of a 
compact object, is of the order of a few
ten seconds \citep{ignazio}, and it gives the typical duration of the GRB in our model. 
In Table 1 we give the measured duration and the estimated electromagnetic energy 
(assuming isotropic emission) 
of the GRBs associated with Fe emission or absorption lines.
All bursts last at least 10 s. 
According to our model
the firsts few ten seconds correspond to a prompt $\gamma$-rays emission, while
the subsequent emission should be interpreted as the beginning of the afterglow.
Actually it has been found that at least the second half of the prompt emission
of long bursts is likely due to afterglow \citep{frontera}.
We would like to remark however that we are not suggesting that all the GRBs
should be explained in our model. In particular long and energetic bursts could be
originated e.g. by collapsars \citep{woosley}. On the other hand,
the variety of GRB durations could be explained within the QS formation scenario itself
making use of the ``unstable photon decay'' mechanism proposed by \citet {ouyedsannino}. 

Next we consider the total energy $\Delta E$ released in the transition 
from a metastable HS (with hyperonic degrees of freedom) 
to HyS or QS
(which final state is reached in this transition depends on 
the details of the QM EOS and in particular on the value of 
the bag constant). 
The energy released is calculated as the difference 
between the gravitational mass of the metastable HS and 
that of the final stable HyS (or QS) having the same baryonic mass \citep{bd00}. 
In Table 2 we report the energy released for various values of the
bag constant $B$ and of the surface tension $\sigma$. Notice that 
the transition will take place when the nucleation time will be 
reduced to a value of the
order years, due e.g. to mass accretion on the HS 
(recall the exponential dependence of the nucleation time
on the mass of the HS, as shown in Fig. 1). Therefore the total energy
released in the collapse will be always of the same order of magnitude,
once the parameters of the model have been fixed.
As shown in  Table 2, the released energy is in the range
$(3 - 5)\times 10^{52}$ ergs for all the sensible choices of the EOS parameters.
The ``critical mass'' $M_{\mathrm{cr}}$ of a metastable HS  having a lifetime 
$\tau = 1$~yr is in the range $(0.9 - 1.4)~ M_\odot$. 
When the mass of the HS reaches a value near $M_{\mathrm{cr}}$, the conversion 
process takes place. 
It is worth mentioning that the energy released in the conversion can be larger 
if a diquark condensate forms inside QM, see e.g. \citet{sannino}.

To generate a strong GRB, an efficient mechanism to transfer the
energy released in the collapse into an electron-photon plasma is needed.
In an earlier work \citep{fryer1998} it was this difficulty that
hampered the possibility to connect GRBs and the hadronic-quark matter
phase transition in compact stars. Only more recently it was noticed \citep{salmonson99}
that near the surface of a compact stellar object, due to
general relativity effects,
the efficiency of the neutrino-antineutrino annihilation into 
$e^+ e^-$ pairs is strongly enhanced with respect to the Newtonion case.
The efficiency of the conversion of neutrinos in $e^+ e^-$ pairs
could be as high as 10$\%$.
In the computation of the energy associated with the
final GRB we must take into account
the possibility of a moderate anisotropy of the electron motion,
due to the presence of the magnetic field of the star,
which will in turn generate a moderate anisotropy of the burst emission 
\footnote{Dramatic effects of a time-dependent magnetic field have been
discussed e.g. by \citet{kluzniak}.}.
Other anisotropies in the GRB emission could be generated by the rotation of the 
star which could affect the efficiency
of the neutrino-antineutrino annihilation due to general relativity effects. 
On the basis of these considerations, 
the energy deposited in the burst could be sufficient to explain
the isotropic energy of the GRBs listed in Table 1.
We must also recall that more efficient ways to generate photons
and/or $e^+ e^-$ pairs have been proposed in the literature, based
on the decay of axion-like particles \citep{noi2000}. This mechanism
would have an extremely high efficiency and would transfer
most of the energy produced in the collapse into GRB electromagnetic energy.

There are various specific signatures of the mechanism we are suggesting.
First, two classes of stars having similar masses but rather different radii should
exist: a) pure (metastable) HS, with radii in the range 12--20 km,
as is the case of the compact star 1E 1207.4-5209, assuming $M=1.4M_\odot$ \citep{sanwal},
and b) HyS or QS with radii in the range 6--8 km \citep{bomb97,li99a,li99b,dra02}.
Second, all the GRBs generated by the present mechanism should
have approximately the same isotropic energy and a duration of at least 10 s.

\section{Conclusions}
We propose the following origin for at least
some of the GRBs having a duration of tens of seconds. They
can be associated with the transition from a metastable HS
to a more compact HyS or a QS. The time delay between the supernova
explosion originating the metastable HS and the GRB 
is regulated by the process of matter accretion on the HS.
While most of the stellar objects 
obtained by a SN explosion will possibly have a mass larger than
$M_{\mathrm{cr}}$ and will therefore directly
stabilize as HyS or QS at the moment of the SN explosion, 
in a few cases the mass of the protoneutron
star will be low enough not to allow the immediate production
of QM inside the star. Only when the star will acquire
enough mass, the process of QM formation could take place.
Due to the surface tension between
the hadronic matter and the QM the star will become metastable.
The later collapse into a stable HyS or QS will generate a powerfull GRB.
It can be interesting to notice that, in order to have a not too small value
for $M_{\mathrm{cr}}$, a relatively large value for the bag constant $B$ has to be
choosen, $B^{1/4}\sim$ 170 MeV, which turns out to be the prefered value in many
hadronic physics calculations (see e.g. \citep{thomas}). 
In this situation the final state is an HyS and not a QS.

\bigskip
\bigskip

It is a pleasure to thank Elena Pian and Luciano Rezzolla for very useful discussions.


\newpage

\begin{figure}
\epsscale{0.8}
\plotone{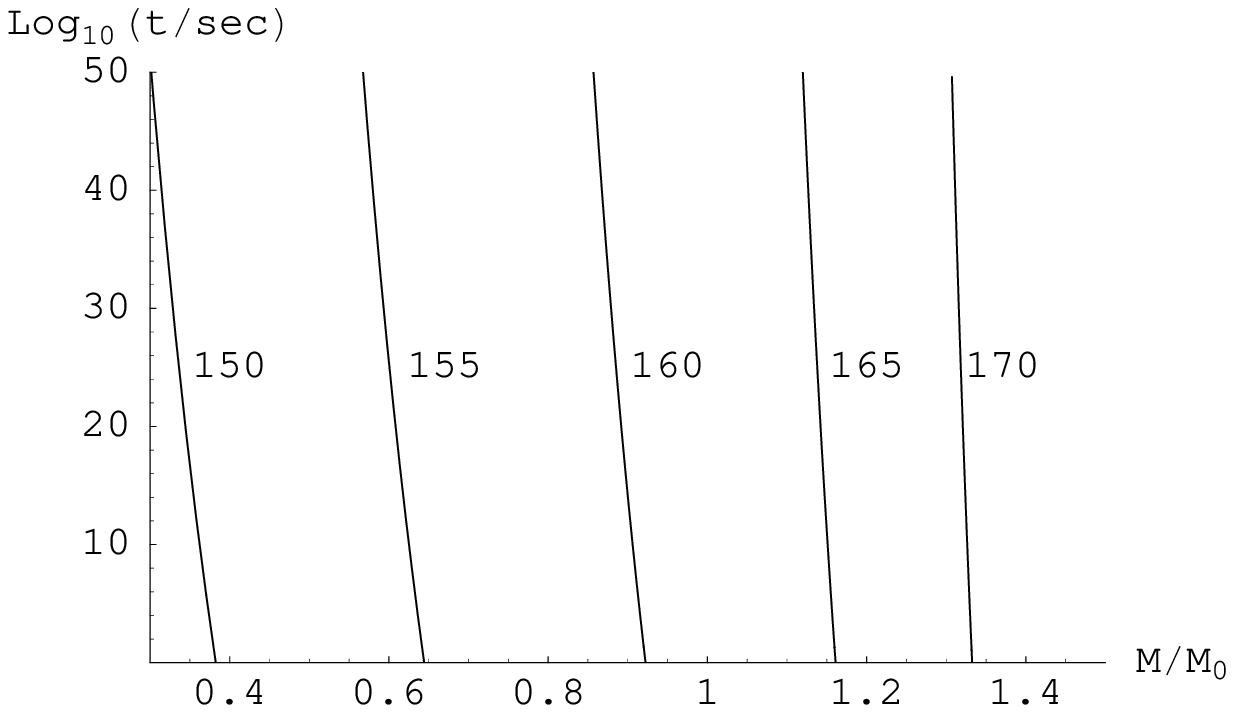}
\caption{Time needed to form a quark matter droplet as a function of 
the mass of the HS for five different values of $B^{1/4}$[MeV].
The hadronic phase has the GM3 parameters set 
with hyperons, the quark phase has $m_s=150$ MeV and a surface
tension $\sigma=30$ MeV/fm$^2$ is assumed. }
\end{figure}

\bigskip
\bigskip
\bigskip

\begin{table}
\begin{center}

\begin{tabular}{|c|c|c|} \hline
GRB      &     duration [s]     &    E$_\mathrm{iso}$/10$^{51}$ erg \\[0.5ex]
\hline
\hline
970508  &  20  & 7  \\
970828  &  160 & 270 \\
990705  &  42  & 210 \\
991216  &  20  & 500 \\
000214  &  10  & 9  \\
011211  &  270 & 50 \\
\hline
\end{tabular}
\caption{Duration and energy released (assuming isotropy) of the GRB
associated with the presence of emission or absorption Fe lines in the spectrum.
The data have been estracted from \citet{Amati02} and \citet{bloom}.}

\end{center}
\end{table}

\begin{table}
\begin{center}

\begin{tabular}{|c|c|c|c|} \hline
B$^{1/4}$[MeV]   &  $\sigma$[MeV/fm$^2$]  & $M_{cr}/M_\odot$  & 
$\Delta$ E [10$^{51}$ erg]    \\[0.5ex]
\hline
\hline
170  &  20   &   1.25   &  30.0  \\
170  &  30   &   1.33   &  33.5  \\
170  &  40   &   1.39   &  38.0  \\
165  &  30   &   1.15   &  38.6  \\
160  &  30   &   0.91   &  45.7  \\
\hline
\end{tabular}
\caption{Critical mass $M_{cr}$ of the metastable hadronic star  
(in unit of the mass of the sun $M_{\odot} = 1.989\times 10^{33}$~g)   
and energy released $\Delta E$ in the conversion to hybrid star
assuming the hadronic star mean life time $\tau$ equal to 1 year.  
Results are reported for various choices of the surface tension $\sigma$ 
and of the bag constant $B$. The strange quark mass is taken equal 
to 150 MeV. For the hadronic matter EOS the GM3 model with hyperons \citep{gm} 
has been used.} 

\end{center}
\end{table}


\begin{thebibliography}{}
\bibitem[Amati et al.(2000)]{Amati00}
Amati, L., et al. 2000, Science, 290, 953.
%
\bibitem[Amati et al.(2002)]{Amati02}
Amati, L., et al. 2002, A\&A, 390,81.
%
\bibitem[Antonelli et al.(2000)]{Antonelli00}
Antonelli, L.A., et al. 2000, ApJ, 545, L39. 
%
\bibitem[Berezhiani \& Drago(2000)]{noi2000}
Berezhiani, Z., \& Drago, A. 2000, Phys. Lett. B, 473, 281.
%
%
\bibitem[Bloom et al.(2001)]{bloom}
Bloom, J.S., Frail, D.A., \& Sari, R. 2001, Astronomical J., 121, 2879.
%
\bibitem[Bombaci(1997)]{bomb97}
Bombaci, I. 1997, Phys. Rev. C,  55, 1587.
%
\bibitem[Bombaci \& Datta(2000)]{bd00}
Bombaci, I., \& Datta, B. 2000, ApJ, 530, L69.
%
\bibitem[B\"ottcher et al.(2002)]{Bottcher02}
B\"ottcher, M., Fryer, C.L., \& Dermer, C.D. 2002, ApJ, 567, 441
%
%
\bibitem[Cheng \& Dai(1996)]{cd96}
Cheng, K.S., \& Dai, Z.G. 1996, Phys. Rev. Lett., 77, 1210. 
%
\bibitem[Cheng et al.(1998)]{che98}
Cheng, K.S., Dai, Z.G., Wei, D.M., \& Lu, T. 1998, Science, 280, 407.
%
\bibitem[Drago \& Lavagno(2001)]{andrea}
Drago, A. \& Lavagno, A. 2001, Phys. Lett. B, 511, 229.
%
\bibitem[Drake et al.(2002)]{dra02}
Drake, J.J., et al. 2002, ApJ, 572, 996.
%
%
\bibitem[Frontera et al.(2000)]{frontera}
Frontera, F., et al. 2000, ApJs, 127, 59.
%
\bibitem[Fryer \& Woosley(1998)]{fryer1998}
Fryer, C.L., \& Woosley, S.E. 1998, ApJ, 501, 780.
%
\bibitem[Glendenning \& Moszkowski(1991)]{gm}
Glendenning, N.K., \& Moszkowski, S.A. 1991, Phys. Rev. Lett., 67, 2414.
%
\bibitem[Glendenning(2000)]{glenbook}
Glendenning, N.K. 2000, Compact Stars, 2nd edition (Springer Verlag).
%
\bibitem[Heiselberg et al.(1993)]{hei}
Heiselberg, H., Pethick, C.P. \& Staubo, E.F. 1993, Phys. Rev. Lett., 70, 1355. 
%
\bibitem[Heiselberg \& Hjorth-Jensen(2000)]{martino}
Heiselberg, H. \& Hjorth-Jensen, M. 2000, Phys. Rept, 328, 237.
%
\bibitem[Hong et al.(2001)]{sannino}
Hong, D.K., Hsu, S.D.H., \& Sannino, F. 2001, Phys. Lett. B., 516, 362.
%
\bibitem[Horvath \& Benvenuto(1988)]{HB88}
Horvath, J.E. \& Benvenuto, O.G. 1988, Phys. Lett. B, 213, 516.
%
\bibitem[Iida \& Sato(1998)]{iida}
Iida, K., \& Sato, K. 1998, Phys. Rev. C, 58, 2538. 
%
\bibitem[Kluzniak \& Ruderman(1998)]{kluzniak}
Kluzniak, W., \& Ruderman, M. 1998, ApJ, 505, L113.
%
\bibitem[Lazzati et al.(2001)]{Lazzati01}
Lazzati, D., et al. 2001, ApJ, 556, 471
%
\bibitem[Li et al.(1999a)]{li99a}
Li, X.D., Bombaci, I., Dey, M., Dey, J. \& van den Heuvel, E.P.J. 1999, 
Phys. Rev. Lett., 83, 3776.
%
\bibitem[Li et al.(1999b)]{li99b}
Li, X.D., Ray, S., Dey, J., Dey, M. \& Bombaci, I. 1999, ApJ, 527, L51. 
%
\bibitem[Lifshitz \& Kagan(1972)]{lif}
Lifshitz, I.M., \& Kagan, Yu. 1972, Zh. Eksp. Teor. Fiz., 62, 385 
[Sov. Phys. JETP, 35, 206].
%
\bibitem[MacFadyen \& Woosley(1999)]{woosley}
MacFadyen, A., \& Woosley, S.E. 1999, ApJ, 524, 262.
%
\bibitem[Masden(1993)]{madsen}
Madsen, J. 1993, Phys. Rev. D, 47, 5156.
%
\bibitem[M\'esz\'aros \& Rees(2001)]{Meszaros01}
M\'esz\'aros, P., \& Rees, M.J. 2001, ApJ, 556, L37.
%
%
\bibitem[Olinto(1987)]{Oli87}
Olinto, A. 1987, Phys. Lett. B, 192, 71.
%
\bibitem[Ouyed et al.(2002)]{ouyed}
Ouyed, R., Dey, J., \& Dey, M. 2002, A\&A, 390, L39.
%
\bibitem[Ouyed \& Sannino(2002)]{ouyedsannino}
Ouyed, R., \& Sannino, F. 2002, A\&A, 387, 725.
%
\bibitem[Piro et al.(1999)]{Piro99}
Piro, L., et al. 1999, ApJ, 514, L73
%
\bibitem[Piro et al.(2000)]{Piro00}
Piro, L., et al. 2000, Science, 290, 955
%
\bibitem[Prakash et al.(1997)]{ignazio}
Prakash, M., Bombaci, I., Prakash, M., Ellis, P.J., Lattimer, J.M. \& Knorren, R.
1997, Phys. Rept., 280, 1.
%
\bibitem[Rees \& M\'esz\'aros(2000)]{Rees00}
Rees, M.J., \& M\'esz\'aros, P. 2000, ApJ, 545, L73.
%
\bibitem[Reeves et al.(2002)]{Reeves02}
Reeves, J.N., et al. 2002, Nature, 414, 512.
%
\bibitem[Salmonson \& Wilson(1999)]{salmonson99}
Salmonson, J.D., \& Wilson, J.R. 1999, ApJ, 517, 859.
%
%
\bibitem[Sanwal et al.(2002)]{sanwal}
Sanwal, D., Pavlov, G.G., Zavlin, V.E., \& Teter, M.A. 2002, ApJ, 574, L61.
%
\bibitem[Steffens et al.(1995)]{thomas}
Steffens, F.M., Holtmann, H., \& Thomas, A.W. 1995, Phys. Lett. B., 358, 139.
%
%
\bibitem[Vietri \& Stella(1998)]{Vietri98}
Vietri, M., \& Stella, L. 1998, ApJ, 507, L45.
%
\bibitem[Wang et al.(2000)]{wang}
Wang, X.Y., Dai, Z.G., Lu, T., Wei, D.M., \& Huang, Y.F. 2000, A\&A, 357, 543.
%
\bibitem[Xu(2002)]{xu02}
Xu, R.X. 2002, ApJ, 570, L65. 
%
\bibitem[Yoshida et al.(1999)]{Yoshida99}
Yoshida, A., et al. 1999, A\&A Suppl., 138, 433
%
\end{thebibliography}
\end{document}